\documentclass{PoS}

\usepackage{amsmath}
\usepackage{amssymb}
\usepackage{graphicx}
 \usepackage{xcolor}
\usepackage[T1]{fontenc} 
\usepackage{slashed}
\usepackage{xspace}
\usepackage{multirow}

\usepackage[utf8]{inputenc}

\newcommand\SARAH{{\tt SARAH}\xspace}

\newcommand\FeynArts{{\tt FeynArts}\xspace}

\newcommand\FlavorKit{{\tt FlavorKit}\xspace}
\newcommand\FormCalc{{\tt FormCalc}\xspace}
\newcommand\CalcHep{{\tt CalcHep}\xspace}
\newcommand\CompHep{{\tt CompHep}\xspace}

\newcommand\WHIZARD{{\tt WHIZARD}\xspace}
\newcommand\OMEGA{{\tt O'Mega}\xspace}

\newcommand\SPheno{{\tt SPheno}\xspace}

\newcommand\Vevacious{{\tt Vevacious}\xspace}

\newcommand\Mathematica{{\tt Mathematica}\xspace}

\newcommand{\Fortran}{\texttt{Fortran}\xspace}

\title{Tutorial to SARAH}

\ShortTitle{Tutorial to SARAH}

\author{\speaker{Florian Staub}\\
       Theoretical Physics Department, CERN, Geneva, Switzerland\\
        E-mail: \email{florian.staub@cern.ch}}

\abstract{I give in this brief tutorial a short practical introduction to the \Mathematica package \SARAH. First, it is shown how an existing model 
file can be changed to implement a new model in \SARAH. In the second part, masses, vertices and renormalisation group equations are calculated 
with \SARAH. Finally, the main commands to generate model files and output for other tools are summarised. }

\FullConference{Proceedings of the Corfu Summer Institute 2015 "School and Workshops on Elementary Particle Physics and Gravity"\\
		1-27 September 2015\\
		Corfu, Greece}

\begin{document}

\section{Introduction}
The \Mathematica package \SARAH \cite{Staub:2008uz,Staub:2009bi,Staub:2010jh,Staub:2012pb,Staub:2013tta,Staub:2015kfa} was created to simplify the study of new models: \SARAH has been optimized for an easy, fast and exhaustive study of non-minimal SUSY and non-SUSY models. It can calculate all important tree-level properties of the model, i.e.  mass matrices and vertices, and interfacing this information to Monte-Carlo (MC) tools, but calculates also one-loop self-energies and two-loop  renormalization group equations (RGEs). Using this information, \SARAH became the first 'spectrum-generator-generator': all analytical information derived by \SARAH can be exported to \Fortran code which provides a fully-fledged spectrum generator based on \SPheno \cite{Porod:2003um,Porod:2011nf}. A \SPheno version created in that way calculates a large number of flavour observables using the \FlavorKit \cite{Porod:2014xia}, and provides even Higgs masses at the two-loop level for a given model \cite{Goodsell:2014bna,Goodsell:2015ira}. An detailed overview to these calculations has been given in the lecture at this school~\cite{Staub:2015iza}, and many more details about the framework are given in Ref.~\cite{Staub:2015kfa}. I focus now on a short practical introduction to \SARAH. \\
First, I briefly demonstrate how an existing model in \SARAH can quickly be changed to a new model, and how the model file is used for basic calculations. I assume in the following that \SARAH is located in the directory {\tt [\$SARAH]}. \\
The simple example which is considered in the following is to extent the superpotential of the NMSSM to include dimensionful terms 
\begin{equation}
W = \dots + \mu \hat H_d \hat H_u + \frac12 \mu_s \hat S \hat S 
\end{equation}
Afterwards, I show how the new model can be used and how information about the model is calculated with \SARAH. 

\section{Changing the model file}
To get started we copy the model directory of the NMSSM to a new subdirectory called {\tt CorfuNMSSM} and rename the file {\tt NMSSM.m} in this directory to {\tt CorfuNMSSM.m}:
\begin{verbatim}
> cp -R [\$SARAH]/Models/NMSSM   [\$SARAH]/Models/CorfuNMSSM 
> mv [\$SARAH]/Models/CorfuNMSSM/NMSSM.m   
                     [\$SARAH]/Models/CorfuNMSSM/CorfuNMSSM.m 
\end{verbatim}
The next step is to add the new terms to the superpotential. For this purpose, we open {\tt CorfuNMSSM.m} in a text editor and change the line which defines the superpotential to 
\begin{verbatim}
SuperPotential = Yu u.q.Hu - Yd d.q.Hd - Ye e.l.Hd + \
           \[Lambda] Hu.Hd.s + \[Kappa]/3 s.s.s \
                 + Mu Hu.Hd + 1/2 MuS s.s;
\end{verbatim}
In order to have a better looking \LaTeX\ output in the following, we slightly extend {\tt parameters.m} also included in the directory {\tt CorfuNMSSM}:
\begin{verbatim}
ParameterDefinitions = { 

{Mu, {LaTeX -> "\\mu"}},
{B[Mu], {LaTeX -> "B_\\mu"}},
{MuS, {LaTeX -> "\\mu_S"}},
{B[MuS], {LaTeX -> "B_S"}},

{g1,        { Description -> "Hypercharge-Coupling"}}, 
...
\end{verbatim}
We are already done and we can run the model now. 

\section{Running \SARAH}
We open \Mathematica and run the new model via 
\begin{verbatim}
<<[\$SARAH]/SARAH.m;
Start["CorfuNMSSM"];
\end{verbatim}
We find the expected warnings from \SARAH that the defined $Z_3$ symmetry is not respected in the new terms:
\begin{verbatim}
Superpotential::ViolationGlobal: Warning! 
   Global symmetry Z3 not conserved in {Hu,Hd}
Superpotential::ViolationGlobal: Warning! 
   Global symmetry Z3 not conserved in {s,s} 
\end{verbatim}
However, we can ignore this warning and continue with the study. After a few seconds, \SARAH is finished with the initialization of the model. 

\section{Calculating analytical properties}
\subsection{Tree-level properties and loop corrections}
After the initialization of the model, we can for instance check that the new terms appear in all mass matrices. For this purpose, we run 
\begin{verbatim}
 MassMatrix[Chi]
\end{verbatim}
and get the neutralino mass matrix which reads
\begin{equation}
\left(
\begin{array}{ccccc}
 \text{MassB} & 0 & -\frac{\text{g1} \text{vd}}{2} &
   \frac{\text{g1} \text{vu}}{2} & 0 \\
 0 & \text{MassWB} & \frac{\text{g2} \text{vd}}{2} &
   -\frac{\text{g2} \text{vu}}{2} & 0 \\
 -\frac{\text{g1} \text{vd}}{2} & \frac{\text{g2} \text{vd}}{2} & 0
   & -\text{Mu}-\frac{\lambda  \text{vS}}{\sqrt{2}} &
   -\frac{\lambda  \text{vu}}{\sqrt{2}} \\
 \frac{\text{g1} \text{vu}}{2} & -\frac{\text{g2} \text{vu}}{2} &
   -\text{Mu}-\frac{\lambda  \text{vS}}{\sqrt{2}} & 0 &
   -\frac{\lambda  \text{vd}}{\sqrt{2}} \\
 0 & 0 & -\frac{\lambda  \text{vu}}{\sqrt{2}} & -\frac{\lambda 
   \text{vd}}{\sqrt{2}} & \text{MuS}+\sqrt{2} \kappa  \text{vS} \\
\end{array}
\right)
\end{equation}
In the same way, we can check the Higgs mass matrices {\tt MassMatrix[hh]} and {\tt MassMatrix[Ah]} and find that there not only {\tt MuS} and {\tt Mu} shows up, but also the automatically generated soft-breaking terms {\tt B[MuS]} and {\tt B[Mu]}. \\

The minimisation conditions of the model this model are returned via 
\begin{verbatim}
TadpoleEquations[EWSB] 
\end{verbatim}
Again, one sees that the news terms are consistently taken into account everywhere. As a simple check that gauge invariance works, we can solve the tadpole equations of the soft-breaking masses $m_{H_d}^2$, $m_{H_u}^2$ and $m_S^2$, plug these solution into mass matrix of the pseudo-scalars and check the lightest eigenvalue for the case of real parameters ({\tt conj[x\_]->x}) and Landau gauge ({\tt RXi[\_]->0}):\\
\begin{verbatim}
Solve[TadpoleEquations[EWSB] == 0, {mHd2, mHu2, ms2}][[1]];
MassMatrix[Ah] /. % /. {RXi[_] -> 0, conj[x_] -> 0};
Eigenvalues[%]
\end{verbatim}
As expected, the first eigenvalue is exactly zero. \\

We can now either calculate specific vertices like for instance the Higgs-down-quark vertex via
\begin{verbatim}
Vertex[{bar[Fu],Fu,hh}] 
\end{verbatim}
and get the expected result 
\begin{eqnarray*}
&&\Big\{\{\text{bar}(\text{Fu}(\{\text{gt1},\text{ct1}\})),\text{Fu}
   (\{\text{gt2},\text{ct2}\}),\text{hh}(\{\text{gt3}\})\}, \\
&&   \left\{-
   \frac{i \text{ZH}(\text{gt3},2)
   \text{Delta}(\text{ct1},\text{ct2})
   \text{sum}(\text{j2},1,3,\text{conj}(\text{ZUL}(\text{gt2},\text
   {j2})) \text{sum}(\text{j1},1,3,\text{Yu}(\text{j1},\text{j2})
   \text{conj}(\text{ZUR}(\text{gt1},\text{j1}))))}{\sqrt{2}},\text
   {PL}\right\}, \\
&&   \left\{-\frac{i \text{ZH}(\text{gt3},2)
   \text{Delta}(\text{ct1},\text{ct2})
   \text{sum}(\text{j2},1,3,\text{ZUL}(\text{gt1},\text{j2})
   \text{sum}(\text{j1},1,3,\text{ZUR}(\text{gt2},\text{j1})
   \text{conj}(\text{Yu}(\text{j1},\text{j2}))))}{\sqrt{2}},\text{P
   R}\right\}\Big\} 
\end{eqnarray*}
or we can calculate all vertices at once. For this purpose we run\footnote{Some warnings will appear here and in the following because of missing definitions for $\mu$, $\mu_S$, $B_S$ and $B_\mu$. The reason is that, in order to save time, we did not put all information for these new parameters in {\tt parameters.m} which are possible. However, these warnings can be ignored for many outputs.} 
\begin{verbatim}
ModelOutput[EWSB, IncludeLoopCorrections -> True, 
 VerticesForLoops -> True,WriteTeX -> True]
\end{verbatim}
Here, we calculate not only  the vertices for the eigenstates after electroweak symmetry breaking ({\tt EWSB}), but we also  (i) include the vertices needed for the loop calculations, (ii) calculate the one-loop corrections, and (iii) generate \LaTeX\ files with all information. After 1--2 minutes \SARAH is finished. We can now check for instance all cubic and quartic vertices which are saved in the arrays 
\begin{verbatim}
SA`VertexList[SSS]
SA`VertexList[SSSS]
\end{verbatim}
However, this is a bit lengthy and difficult to read. Therefore, we check the \LaTeX\ output. For this purpose we enter the output directory and make the shell script to run {\tt pdflatex} executable
\begin{verbatim}
cd [\$SARAH]\Output/CorfuNMSSM/EWSB/TeX/
chmod 775 MakePDF.sh
./MakePDF.sh
\end{verbatim}
This script does not only run {\tt pdflatex} but generates also all Feynman diagrams. After some seconds we get a pdf file with 122 pages which contains all information about the model which \SARAH has derived so far. 

\subsection{RGEs}
The next step is to check the RGEs for this model. To do this, we just have to run 
\begin{verbatim}
CalcRGEs[]; 
\end{verbatim}
After not even one minute, \SARAH is finished with the calculation of all two-loop $\beta$-functions. The calculations of the RGEs in \SARAH 
are based on the generic results of Refs.~\cite{Machacek:1983tz,Machacek:1983fi,Machacek:1984zw,Martin:1993zk,Luo:2002ti,Fonseca:2011vn,Goodsell:2012fm,Sperling:2013eva,Sperling:2013xqa,Fonseca:2013bua}.
We can check the RGEs for the new superpotential terms via
\begin{verbatim}
BetaMuij
\end{verbatim}
and get 
\begin{verbatim}
{{Mu, 
  -((3 g1^2 Mu)/5) - 3 g2^2 Mu + 2 Mu \[Lambda] conj[\[Lambda]] + 
     3 Mu trace[Yd, Adj[Yd]] + Mu trace[Ye, Adj[Ye]] + 
     3 Mu trace[Yu, Adj[Yu]], 
  1/50 Mu (207 g1^4 + 90 g1^2 g2^2 + 375 g2^4 - 
     200 \[Kappa] \[Lambda] conj[\[Kappa]] conj[\[Lambda]] - 
     300 \[Lambda]^2 conj[\[Lambda]]^2 - 20 g1^2 trace[Yd, Adj[Yd]] + 
     800 g3^2 trace[Yd, Adj[Yd]] + 60 g1^2 trace[Ye, Adj[Ye]] + 
     40 g1^2 trace[Yu, Adj[Yu]] + 800 g3^2 trace[Yu, Adj[Yu]] - 
     50 \[Lambda] conj[\[Lambda]] (3 trace[Yd, Adj[Yd]] + 
        trace[Ye, Adj[Ye]] + 3 trace[Yu, Adj[Yu]]) - 
     450 trace[Yd, Adj[Yd], Yd, Adj[Yd]] - 
     300 trace[Yd, Adj[Yu], Yu, Adj[Yd]] - 
     150 trace[Ye, Adj[Ye], Ye, Adj[Ye]] - 
     450 trace[Yu, Adj[Yu], Yu, Adj[Yu]])}, 
{MuS, 
   4 MuS (\[Kappa] conj[\[Kappa]] + \[Lambda] conj[\[Lambda]]), 
 -(4/5) MuS (20 \[Kappa]^2 conj[\[Kappa]]^2 + 
     20 \[Kappa] \[Lambda] conj[\[Kappa]] conj[\[Lambda]] + \[Lambda] \
     conj[\[Lambda]] (-3 g1^2 - 15 g2^2 + 10 \[Lambda] conj[\[Lambda]] + 
     15 trace[Yd, Adj[Yd]] + 5 trace[Ye, Adj[Ye]] + 
     15 trace[Yu, Adj[Yu]]))}}
\end{verbatim}
For both parameter the one- and two-loop $\beta$-function is given up to factors $1/16\pi^2$. \\

\SARAH writes also a file to run these RGEs numerically with \Mathematica. One can load this file in any \Mathematica session and use the provided function {\tt RunRGEs} to solve the RGEs numerically. We are going to use this to check if we can reproduce gauge coupling unification:
\begin{verbatim}
<< [\$SARAH]/Output/CorfuNMSSM/RGEs/RunRGEs.m;
sol = RunRGEs[{g1 -> 0.46, g2 -> 0.63, g3 -> 1.05}, 3, 17][[1]];
Plot[{g1[x], g2[x], g3[x]} /. sol, {x, 3, 17}, Frame -> True, Axes -> False];
\end{verbatim}
We first loaded the file and then run the RGEs from $10^3$ to $10^{17}$ GeV. As initial conditions at 1 TeV we used $g_1=0.46$, $g_2=0.63$, $g_3=1.05$. The interpolation function were saved in the variable {\tt sol} which we then used to make a plot. The result is not very surprising:
\begin{center}
\includegraphics{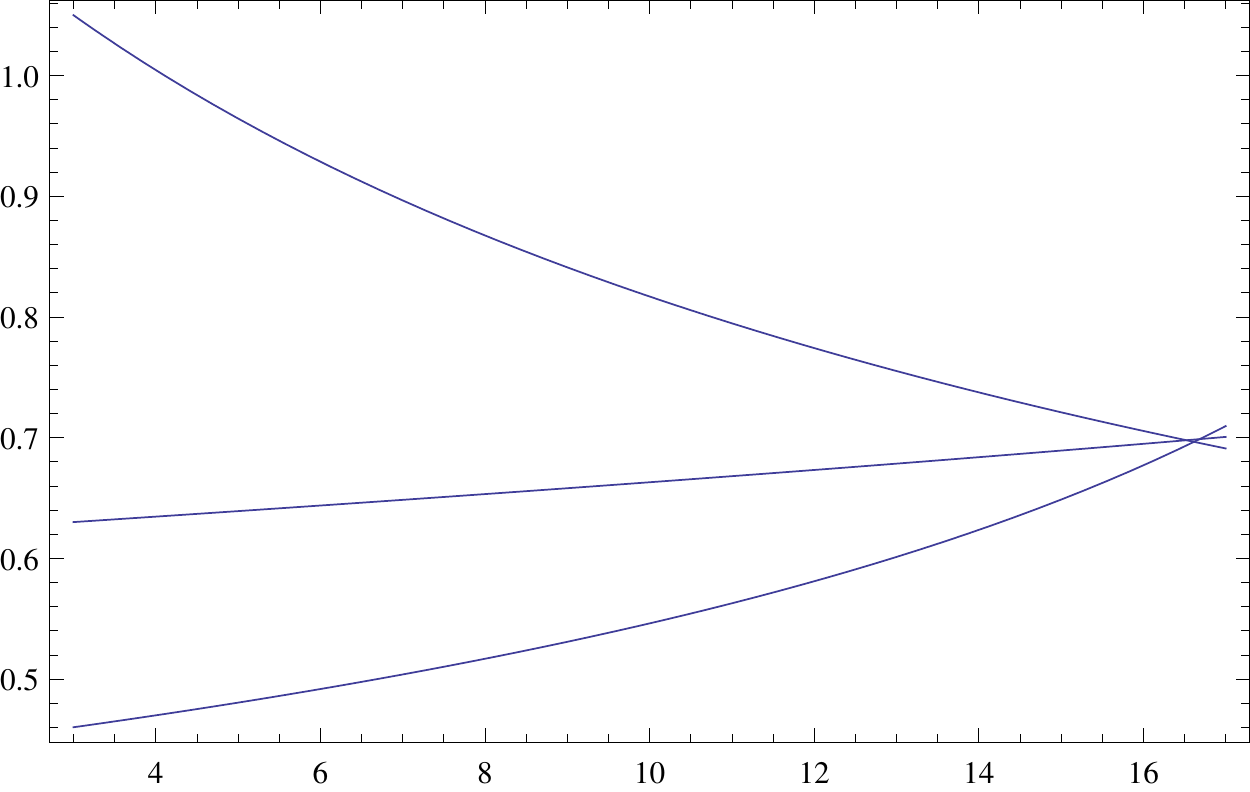}
\end{center}

\section{Outputs for other codes}
Since already all vertices are calculated, we can start now to generate model files for the different tools. The corresponding commands are 
\begin{itemize}
 \item {\tt MakeCHep[]} for \CalcHep/\CompHep  \cite{Pukhov:2004ca,Boos:1994xb}
 \item {\tt MakeFeynArts[]} for \FeynArts/\FormCalc  \cite{Hahn:2000kx,Hahn:1998yk}
 \item {\tt MakeUFO[]} for UFO  \cite{Degrande:2011ua}
 \item {\tt MakeWHIZARD[]} for \WHIZARD/\OMEGA \cite{Kilian:2007gr}
 \item {\tt MakeVevacious[]} for \Vevacious \cite{Camargo-Molina:2013qva}
\end{itemize}
Before we can use 
\begin{verbatim}
 MakeSPheno[]
\end{verbatim}
to generate the Fortran code to implement this model in \SPheno, we would have slightly to modify {\tt [\$SARAH]/Models/CorfuNMSSM/SPheno.m} to define 
the boundary conditions for the new parameters. However, this is beyond the scope of this short tutorial and I refer interested user to Ref.~\cite{Staub:2015kfa} for detailed discussions and examples how to use these commands.

\section{Summary}
I have shown, how a new SUSY singlet extensions can be implemented easily \SARAH starting from the default MSSM model file. I then described how analytical and numerical results can quickly be obtained for this model. 

\section*{Acknowledgements}
I thank the organizers of ``Summer School and Workshop on the Standard Model and Beyond'', Corfu2015, for the invitation and the hospitality during the stay.


\begin{thebibliography}{99}

\bibitem{Staub:2008uz}
  F.~Staub,
  arXiv:0806.0538 [hep-ph].
\bibitem{Staub:2009bi}
  F.~Staub,
  Comput.\ Phys.\ Commun.\  {\bf 181} (2010) 1077
  doi:10.1016/j.cpc.2010.01.011
  [arXiv:0909.2863 [hep-ph]].
\bibitem{Staub:2010jh}
  F.~Staub,
  Comput.\ Phys.\ Commun.\  {\bf 182} (2011) 808
  doi:10.1016/j.cpc.2010.11.030
  [arXiv:1002.0840 [hep-ph]].
\bibitem{Staub:2012pb}
  F.~Staub,
  Comput.\ Phys.\ Commun.\  {\bf 184} (2013) 1792
  doi:10.1016/j.cpc.2013.02.019
  [arXiv:1207.0906 [hep-ph]].
\bibitem{Staub:2013tta}
  F.~Staub,
  Comput.\ Phys.\ Commun.\  {\bf 185} (2014) 1773
  doi:10.1016/j.cpc.2014.02.018
  [arXiv:1309.7223 [hep-ph]].
\bibitem{Staub:2015kfa}
  F.~Staub,
  Adv.\ High Energy Phys.\  {\bf 2015} (2015) 840780
  doi:10.1155/2015/840780
  [arXiv:1503.04200 [hep-ph]].
\bibitem{Porod:2003um}
  W.~Porod,
  Comput.\ Phys.\ Commun.\  {\bf 153} (2003) 275
  doi:10.1016/S0010-4655(03)00222-4
  [hep-ph/0301101].
\bibitem{Porod:2011nf}
  W.~Porod and F.~Staub,
  Comput.\ Phys.\ Commun.\  {\bf 183} (2012) 2458
  doi:10.1016/j.cpc.2012.05.021
  [arXiv:1104.1573 [hep-ph]].

\bibitem{Porod:2014xia}
  W.~Porod, F.~Staub and A.~Vicente,
  Eur.\ Phys.\ J.\ C {\bf 74} (2014) no.8,  2992
  doi:10.1140/epjc/s10052-014-2992-2
  [arXiv:1405.1434 [hep-ph]].


\bibitem{Goodsell:2014bna}
  M.~D.~Goodsell, K.~Nickel and F.~Staub,
  Eur.\ Phys.\ J.\ C {\bf 75} (2015) no.1,  32
  doi:10.1140/epjc/s10052-014-3247-y
  [arXiv:1411.0675 [hep-ph]].

\bibitem{Goodsell:2015ira}
  M.~Goodsell, K.~Nickel and F.~Staub,
  Eur.\ Phys.\ J.\ C {\bf 75} (2015) no.6,  290
  doi:10.1140/epjc/s10052-015-3494-6
  [arXiv:1503.03098 [hep-ph]].
  
\bibitem{Staub:2015iza}
  F.~Staub,
  arXiv:1509.07061 [hep-ph].
  
\bibitem{Machacek:1983tz}
  M.~E.~Machacek and M.~T.~Vaughn,
  Nucl.\ Phys.\ B {\bf 222} (1983) 83.
  doi:10.1016/0550-3213(83)90610-7
\bibitem{Machacek:1983fi}
  M.~E.~Machacek and M.~T.~Vaughn,
  Nucl.\ Phys.\ B {\bf 236} (1984) 221.
  doi:10.1016/0550-3213(84)90533-9
\bibitem{Machacek:1984zw}
  M.~E.~Machacek and M.~T.~Vaughn,
  Nucl.\ Phys.\ B {\bf 249} (1985) 70.
  doi:10.1016/0550-3213(85)90040-9
\bibitem{Martin:1993zk}
  S.~P.~Martin and M.~T.~Vaughn,
  Phys.\ Rev.\ D {\bf 50} (1994) 2282
   Erratum: [Phys.\ Rev.\ D {\bf 78} (2008) 039903]
  doi:10.1103/PhysRevD.50.2282, 10.1103/PhysRevD.78.039903
  [hep-ph/9311340].
\bibitem{Luo:2002ti}
  M.~x.~Luo, H.~w.~Wang and Y.~Xiao,
  Phys.\ Rev.\ D {\bf 67} (2003) 065019
  doi:10.1103/PhysRevD.67.065019
  [hep-ph/0211440].
\bibitem{Fonseca:2011vn}
  R.~M.~Fonseca, M.~Malinsky, W.~Porod and F.~Staub,
  Nucl.\ Phys.\ B {\bf 854} (2012) 28
  doi:10.1016/j.nuclphysb.2011.08.017
  [arXiv:1107.2670 [hep-ph]].
\bibitem{Goodsell:2012fm}
  M.~D.~Goodsell,
  JHEP {\bf 1301} (2013) 066
  doi:10.1007/JHEP01(2013)066
  [arXiv:1206.6697 [hep-ph]].
\bibitem{Sperling:2013eva}
  M.~Sperling, D.~Stöckinger and A.~Voigt,
  JHEP {\bf 1307} (2013) 132
  doi:10.1007/JHEP07(2013)132
  [arXiv:1305.1548 [hep-ph]].
\bibitem{Fonseca:2013bua}
  R.~M.~Fonseca, M.~Malinský and F.~Staub,
  Phys.\ Lett.\ B {\bf 726} (2013) 882
  doi:10.1016/j.physletb.2013.09.042
  [arXiv:1308.1674 [hep-ph]].
\bibitem{Sperling:2013xqa}
  M.~Sperling, D.~Stöckinger and A.~Voigt,
  JHEP {\bf 1401} (2014) 068
  doi:10.1007/JHEP01(2014)068
  [arXiv:1310.7629 [hep-ph]].
  

\bibitem{Pukhov:2004ca} 
  A.~Pukhov,
  hep-ph/0412191.
  
\bibitem{Boos:1994xb} 
  E.~E.~Boos, M.~N.~Dubinin, V.~A.~Ilyin, A.~E.~Pukhov and V.~I.~Savrin,
  hep-ph/9503280.
  
\bibitem{Hahn:2000kx} 
  T.~Hahn,
  Comput.\ Phys.\ Commun.\  {\bf 140}, 418 (2001)
  doi:10.1016/S0010-4655(01)00290-9
  [hep-ph/0012260].
  
\bibitem{Hahn:1998yk} 
  T.~Hahn and M.~Perez-Victoria,
  Comput.\ Phys.\ Commun.\  {\bf 118}, 153 (1999)
  doi:10.1016/S0010-4655(98)00173-8
  [hep-ph/9807565].
  
\bibitem{Kilian:2007gr} 
  W.~Kilian, T.~Ohl and J.~Reuter,
  Eur.\ Phys.\ J.\ C {\bf 71}, 1742 (2011)
  doi:10.1140/epjc/s10052-011-1742-y
  [arXiv:0708.4233 [hep-ph]].
  
\bibitem{Degrande:2011ua} 
  C.~Degrande, C.~Duhr, B.~Fuks, D.~Grellscheid, O.~Mattelaer and T.~Reiter,
  Comput.\ Phys.\ Commun.\  {\bf 183}, 1201 (2012)
  doi:10.1016/j.cpc.2012.01.022
  [arXiv:1108.2040 [hep-ph]].
  
\bibitem{Camargo-Molina:2013qva} 
  J.~E.~Camargo-Molina, B.~O'Leary, W.~Porod and F.~Staub,
  Eur.\ Phys.\ J.\ C {\bf 73}, no. 10, 2588 (2013)
  doi:10.1140/epjc/s10052-013-2588-2
  [arXiv:1307.1477 [hep-ph]].

\end{thebibliography}
\end{document}